# A Robust Integrated Multi-Strategy Bus Control System via Deep Reinforcement Learning

Qinghui Nie[a], Jishun Ou[a], Haiyang Zhang[a], Jiawei Lu[b], Shen Li[c] and Haotian Shi[d]*

[a]*College of Architectural Science and Engineering, Yangzhou University, Yangzhou, China;*

[b]*H. Milton Stewart School of Industrial and Systems Engineering, Georgia Institute of Technology, Atlanta, Georgia*

[c]*Department of Civil Engineering, Tsinghua University, Beijing, China*

[d]*Department of Civil and Environmental Engineering, University of Wisconsin-Madison, Madison, United States*

*Corresponding author: Haotian Shi (hshi84@wisc.edu)

# A Robust Integrated Multi-Strategy Bus Control System via Deep Reinforcement Learning

An efficient urban bus control system has the potential to significantly reduce travel delays and streamline the allocation of transportation resources, thereby offering enhanced and user-friendly transit services to passengers. However, bus operation efficiency can be impacted by bus bunching. This problem is notably exacerbated when the bus system operates along a signalized corridor with unpredictable travel demand. To mitigate this challenge, we introduce a multi-strategy fusion approach for the longitudinal control of connected and automated buses. The approach is driven by a physics-informed deep reinforcement learning (DRL) algorithm and takes into account a variety of traffic conditions along urban signalized corridors. Taking advantage of connected and autonomous vehicle (CAV) technology, the proposed approach can leverage real-time information regarding bus operating conditions and road traffic environment. By integrating the aforementioned information into the DRL-based bus control framework, our designed physics-informed DRL state fusion approach and reward function efficiently embed prior physics and leverage the merits of equilibrium and consensus concepts from control theory. This integration enables the framework to learn and adapt multiple control strategies to effectively manage complex traffic conditions and fluctuating passenger demands. Three control variables, i.e., dwell time at stops, speed between stations, and signal priority, are formulated to minimize travel duration and ensure bus stability with the aim of avoiding bus bunching. We present simulation results to validate the effectiveness of the proposed approach, underlining its superior performance when subjected to sensitivity analysis, specifically considering factors such as traffic volume, desired speed, and traffic signal conditions.

**1. Introduction**

Bus bunching is a phenomenon where two or more buses on the same route arrive at a bus station in close succession due to uncertainties in bus operation systems. The uncertainties may stem from variations in travel times caused by fluctuating traffic conditions, diverse driver behaviors, as well as time-varying passenger demand. When a bus falls behind the schedule, the headway between it and the preceding bus increases, leading to longer dwell times as passengers accumulate at downstream stops. As the following bus arrives, fewer passengers are present at the stop, resulting in a shorter dwell time. Consequently, the average headway deviation in the bus system becomes substantial (Iliopoulou et al., 2018). The bus bunching problem poses a serious challenge in the design and operation of bus systems.

To alleviate the bus bunching problem, numerous bus control strategies have been proposed. One prevalent method involves implementing control strategies at bus stations, such as bus holding or stop-skipping. The bus holding strategies encompass naïve schedule control and headway control. Within these two strategies, transit agencies introduce schedule slacks to mitigate the bus bunching problem. However, the holding strategies may fall short in addressing localized disruptions within the network (Daganzo, 2009). The stop-skipping strategies are able to alleviate bus bunching to some degree, but easily result in increased travel time and inconvenience for passengers. Another common solution to alleviate bus bunching is interstation control, which includes bus speed control and bus signal priority. An adaptive bus speed control strategy, encouraging cooperative control between buses, emerges as a promising alternative to traditional holding strategies (Daganzo and Pilachowski, 2011). The signal priority strategy is suitable for real-time control since it adjusts the timing of traffic signals in response to bus arrivals. Other strategies for tackling the bus bunching problem include restricting boarding (Zhao et al., 2016) and vehicle substitution (Petit et al., 2019).

Although the traditional methods mentioned above have been successfully applied to mitigate bus bunching, there are still three main issues that limited the applicability of the methods. First, most existing methods develop a single strategy for bus control, cannot fully exploit the complementarity of multiple strategies while overlooking the constraints posed by road and station conditions (Wang and Sun, 2020; Tian et al., 2022). Second, the majority of the methods only consider a single type of uncertainty and assume the disturbance source is deterministic or adheres to an analytical distribution, which is inconsistent with real-world conditions. Third, these methods mainly rely on exact solution algorithms which pose difficulties in locating the optimal solution within a feasible computation time, highlighting a need for more efficient problem-solving techniques in the field of transit management.

The advent of Connected and Automated Vehicles (CAV) technologies provide bus operation systems the ability to real-time access to passenger demand, roadway traffic speed, and signal information and allow decision-makers to integrate multiple control strategies by facilitating more efficient communication and collaboration between buses and infrastructure and exploiting various information sources (Shen et al., 2019). Despite the huge potential of CAV technology, its application and integration to transit buses are limited. Recently, several efforts have been made to develop bus control strategies within a CAV environment (Laskaris et al., 2020; Shi et al., 2022), and demonstrated that a cooperative multi-strategy approach within a CAV environment can yield significant benefits for bus operation systems (Abdelhalim and Abbas, 2021).

As a popular and powerful machine learning paradigm, reinforcement learning provides a means to learn strategies through online interactions with the environment, making it well-suited for practical scenarios such as bus operation control. Deep reinforcement learning (DRL), a combination of reinforcement learning and deep learning, utilizes neural networks to approximate value functions or policies, rendering it especially useful for processing high-dimensional state and action spaces in complex environments. DRL is now extensively employed for controlling multi-agent systems due to its exceptional generalization performance and real-time decision-making capabilities (Farazi et al., 2021).

Based on the CAV technology and DRL algorithms, we propose a robust integrated multi-strategy bus control approach to effectively alleviating bus bunching. Our approach integrates historical bus operation data and real-time detector data to simulate a realistic bus operation environment. The control strategy fusions the bus holding, inter-station speed control, and transit signal adjustment. A distributed proximal policy optimization (DPPO) algorithm is adopted to efficiently train a control strategy that can achieving the desired performance. A series of numerical experiments are conducted to evaluate the proposed approach. Results show that the three single control strategies can cooperate with each other and be integerated to exhibit superior performance even with fluctuating and uncertainty traffic conditions. The main contributions are summarized as follows:

(1) We present a multi-strategy fusion approach based on distributed DRL for dynamic bus operation control. It aims to enhance the longitudinal control of connected and automated buses traversing signalized corridors with respect to varying traffic conditions. This approach amalgamates and implements an array of control strategies, assessing the practicality of each to augment the flexibility of overall control. Particularly, this approach encompasses three interdependent control variables: dwell time at stops, speed between stops, and signal priority. These variables synergistically operate under diverse traffic scenarios to ensure consistent schedule adherence and headway regularity.
(2) We consider the uncertainties inherent in bus operation systems in the modeling process and design a practical bus control environment that integrates spatial-temporal variations in disturbances by leveraging both historical and real-time bus data. Our analysis accommodates two primary sources of uncertainty in realistic bus operations: variability in travel times between stops, and fluctuations in passenger demand at stations. By accounting for these uncertainties and dynamically adjusting bus operations, our approach is capable of significantly mitigating bus bunching and enhancing the overall efficiency of the bus system.

(3) We develop an efficient DPPO-based learning procedure for training the bus control model, demonstrating remarkable solution efficiency and convergence. To achieve multi-agent consensus and ensure bus system stability, we have developed a physics-informed DRL state fusion approach and a tailored reward function, which efficiently integrates the prior physics. Further, we incorporate principles from multi-agent control theory, along with the concept of equilibrium states within bus operations. By utilizing the combined state information of downstream buses, each bus can maintain the equilibrium state, ultimately leading to a stable and effective bus system.

The remainder of the paper is structured as follows: Section 2 discusses the relevant literature. Section 3 presents our DRL-based multi-strategy bus control methodology. In Section 4, a series of numerical examples are provided to illustrate the effectiveness and applicability of the proposed approach. The conclusion is drawn in Section 5.

## 2. Literature review

This section introduces a systematic review on bus control strategies to provide a comprehensive understanding of the related research and identify gaps in the existing literature. In general, according to the spatial configuration of different bus control strategies, existing bus control strategies can be divided into two categories, namely station control and inter-station control (Muñoz et al., 2013).

*2.1 Station control strategies*

Bus holding is a widely used station control strategy to improve the reliability of bus operations. Early studies on bus holding, such as those by Carey (1994), Abkowitz and Lepofsky (1990), and Dessouky et al. (1999), incorporated slack into schedules to effectively reduce bus bunching. However, excessive slack can result in decreased bus service frequency. To address this issue, Zhao et al. (2006) devised a schedule-based analytical model to determine the optimal slack time, aiming to minimize the expected passenger waiting time and set appropriate slack levels. Building upon this, Daganzo (2009) introduced a dynamic method for determining bus holding time at control points using real-time headway information, which further reduced slack. Expanding on this research, Xuan et al. (2011) developed a range of dynamic holding strategies and introduced the concept of virtual timetables, significantly enhancing control efficiency. More recently, He et al. (2020) proposed a target-headway-based holding strategy that adjusts bus operations with the average value of instantaneous headway as the target.

Moreover, holding strategies incorporating real-time predictions of subsequent bus arrival times have emerged as innovative solutions (Berrebi et al., 2018). Bartholdi III and Eisenstein (2012) devised a self-coordinated strategy that does not rely on predetermined schedules or target headways. Instead, it dynamically self-balances headways based on the predicted arrival time of the next bus, offering the potential to further enhance bus service. Berrebi et al. (2018) compared bus holding methods with and without real-time prediction and found that the prediction-based method performed well in maintaining a balance between headway regularity and holding time. However, the implementation of a single transit route improvement strategy may be affected by the interactions between different transit routes in public corridors. To address this issue,

Zhou et al. (2019) developed a coordinated holding strategy based on collaborative control for two bus lines, which exhibited strong performance in simulation results.

Stop-skipping is another widely applied station control strategy used to adjust bus operations. Numerous studies have investigated this strategy to optimize its implementation. Fu et al. (2003) developed a nonlinear integer programming model to determine a stop-hopping strategy, aiming to minimize the total cost for both operators and passengers. Conversely, Sun and Hickman (2005) examined the feasibility of implementing a real-time stop-skipping strategy, allowing passengers to disembark at stops that skipped segments. Due to the uncertainty of passenger boarding and alighting times, the authors proposed a nonlinear integer programming problem, solved using an exhaustive search method. Liu et al. (2013) developed an optimization model for the stop-skipping problem, accounting for the random travel time of buses. The model aimed to minimize the weighted sum of three critical factors: total waiting time, total in-vehicle travel time, and total operating cost. The authors proposed a genetic algorithm as a solution strategy for this problem. Chen et al. (2015) addressed the limitation of bus capacity and the impact of overloading on buses. They proposed a solution to optimize the stop-skipping model using a hybrid artificial bee colony (ABC) algorithm.

Although these station control methods have demonstrated promising outcomes in particular traffic conditions, they also present certain limitations (Ibarra-Rojas et al., 2015; Gkiotsalitis et al., 2021). Specifically, they may cause passenger inconvenience and have a negative impact on their travel experience by increasing the holding time for onboard passengers and preventing passengers from boarding at skipped stops.

*2.2 Inter-station control strategies*

Inter-station control strategies typically involve bus operating speed adjustments and bus signal priority. Daganzo and Pilachowski (2011) proposed an adaptive control method that adjusts the cruising speed of buses in real-time based on anticipated passenger demand information and the distance between leading and following buses. This adaptive control method effectively alleviates bus bunching. Building on this work, Ampountolas and Kring (2021) developed a bus-to-bus cooperative adaptive control strategy that employs bus-following models to regulate speed and headway using nonlinear and linear control rules. Simulation results demonstrate its effectiveness. In a separate study, Liu et al. (2003) proposed a dynamic model that optimizes green time allocation in response to real-time traffic flow conditions, aiming to minimize average delay at intersections. Han et al. (2014) formulated optimal bus priority as a quadratic programming problem, with the objective function being a weighted sum of transit delays and traffic delays. The weights are dynamically adjusted to reflect changing network conditions.

Contrary to the station control strategy, the inter-station control strategy has fewer adverse effects. This is attributed to the utilization of bus operating speed adjustments and signal priority, which can alleviate passenger dissatisfaction by distributing holding time across various segments of the vehicle's trajectory. These strategies complement the inter-station maneuvers within the bus control system, enhancing operational performance. However, the effectiveness of these strategies is constrained by road traffic conditions, and their implementation can prove challenging. Consequently, investigating the integration and application of diverse strategies is crucial.

*2.3 DRL-based control Strategies*

The preceding studies primarily focus on accurately solving constrained stochastic optimization, which can be computationally intensive and challenging to implement. The advent of DRL offers a novel approach to mitigate bus bunching. In the domain of bus control, researchers have proposed various bus coordination control methods. For example, Chen et al. (2016) employed a multi-agent Q-learning algorithm to make coordinated bus holding decisions. Alesiani and Gkiotsalitis (2018) explored a reinforcement learning-based approach for real-time bus holding decision-making, while also considering the impact of holding decisions on other buses. Wang and Sun (2020) proposed a multi-agent DRL framework for developing dynamic holding control strategies, wherein each bus interacts with all other vehicles in the fleet. The learning performance is improved through the application of proximal policy optimization. However, the existing DRL-based bus control framework does not entirely address the non-stationarity issue (Laurent et al., 2011), which leads to higher computational costs.

*2.4 Current status and challenges in the research field*

In conclusion, the existing body of literature has explored numerous bus control strategies aimed at mitigating bus bunching. However, a majority of these studies have not capitalized on the emerging potential of CAV technology, nor have they effectively utilized real-time data or comprehensively addressed uncertainties inherent in bus operations. Additionally, many researchers continue to rely on singular control strategies, thereby constraining the flexibility and practicality of bus control solutions. Traditional optimization methods, which are often employed to tackle these problems, present significant challenges in obtaining successful outcomes. To overcome these limitations, we put forth a distributed DRL-based multi-strategy fusion approach for dynamic bus control. This innovative method harnesses the power of CAV technology to maintain real-time schedule adherence, optimize speed, and prioritize traffic signals, ultimately alleviating the issue of bus bunching.

## 3. Methodology

For the environment setting, we examine a bus transit system characterized by a looped configuration, encompassing $j$ positions that can be categorized into three types: stations, traffic signals, and intra-station roads. This system operates through a cohort of buses traveling unidirectionally along the bus corridor, commencing from station 0 and terminating at station $N$. Each bus is required to make scheduled stops at every station, while all intersections within the system are outfitted with traffic signals. As shown in Figure 1, the environment contains bus stations, roadways, and consecutive signals. For the looped bus system, we make the following assumptions:

(1) Bus capacity is unlimited (Xuan et al., 2011);
(2) Station demand, composed of a nominal value based on historical data and random disturbances within a limited range, is assumed to be known (Li et al., 2019);
(3) Dwell time depends on headway, speed, and passenger demand rate (Li et al., 2019);
(4) Travel time (delay) between bus stations adheres to a normal distribution, informed by historical data (Sánchez-Martínez et al, 2016);
(5) Road traffic conditions allow for speed adjustments in the area between stations (Daganzo and Pilachowski, 2011);
(6) Traffic volume remains constant throughout a loop;
(7) Transit priorities are present at specific intersections (Estrada et al., 2016);
(8) Real-time signal information is available, courtesy of CAV technology (Estrada et al., 2016).

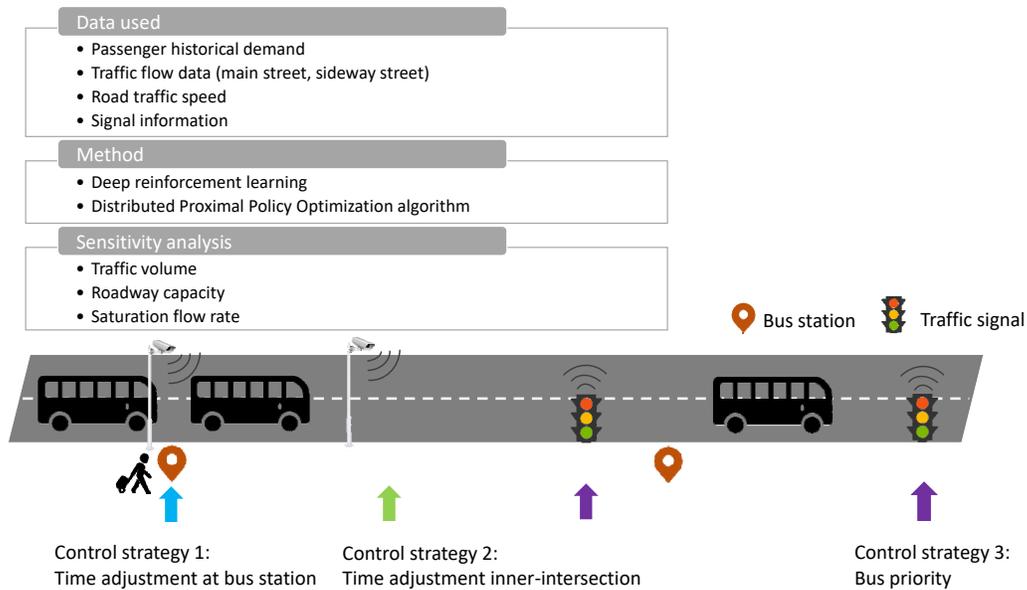

Figure 1. The design of the multi-Strategy Bus Control System.

The methodology can be delineated into three components: data input, DRL environment, and evaluation. During the data input phase, historical passenger demand data and average travel time data between stations are utilized as the primary source of information. Three additional data points are acquired through CAV technology in real time, including traffic flow volume, average road speed, and signal information at upcoming intersections. Deviations in headway and arrival time introduce uncertainties that should be considered within this framework.

The second component of the methodology involves constructing the DRL environment for bus movement control. Planned movements are calculated based on historical passenger demand data, which serves as a target. However, bus operations in complex and dynamic traffic environments can be disrupted by various temporal and spatial disturbances. The study aimed to ensure the authenticity of the environment by accounting for uncertainty factors, including changes in inter-station travel times and fluctuations in station passenger demand. To cope effectively with spatially and temporally fluctuating disturbances, three control forces have been designed to dynamically adjust bus operations by modifying dwell times at stops, speeds between stations, and signal priorities. The process of adjusting controlled values determines the actual movements of the bus. The actual movements are expected to align with the planned movements, and any differences between the two are considered errors that need to be minimized.

The ultimate evaluation phase involves examining the schedule deviation and the headway deviation of bus operation across varying traffic conditions. This analysis provides insight into the alignment between actual and planned movement, thereby allowing for an assessment of the control performance of the proposed method. By scrutinizing these factors, we can gauge the extent to which the proposed approach effectively regulates and optimizes bus operations.

The proposed methodology integrates three strategies, namely the bus holding strategy, cruise speed adjustment strategy, and signal priority strategy, to address the challenges posed by road and bus station conditions. The integration of these strategies offers a comprehensive solution that can be adapted to diverse traffic scenarios. This deviates from conventional research by examining the feasibility of bus control strategies while considering various traffic situations. To address this issue, the study investigates the feasible range for each control type. Control force can be categorized into two types, as outlined by Teng and Jin (2015): instant and accumulative control force. Instant control force refers to control applied at a single point, enabling immediate adoption for the target bus. This can be achieved through strategies such as adjusting dwell time at stations or within inter-station areas. Under varying traffic conditions, including traffic flow volumes, capacity, and saturation flow rates, the feasible range of control forces may differ. A suitable proportion of each control strategy will be optimally trained to ensure peak performance is attained. In certain traffic situations, some strategies may prove ineffective. For instance, in a fully saturated intersection, bus priority may not be activated. The feasibility of control forces will be modeled to demonstrate how these three control strategies collaboratively complement one another under various traffic conditions.

Furthermore, inventory control force represents control that is applied continuously and adopted for the target bus in stages, with varying control quantities. This can be achieved by implementing strategies along the signalized corridor. For instance, traffic signal control along the road can serve as an inventory control force. The cumulative effect of signal priority influences the movement of buses. This study aims to investigate three control strategies that incorporate both immediate and inventory control forces, enhancing the flexibility of control under diverse traffic operation scenarios.

Building on these insights, this paper presents a distributed DRL-based control strategy that factors in traffic volume and signal information to address the bus bunching problem. The approach considers the feasibility of applying diverse control strategies under varying traffic conditions, offering a more comprehensive solution.

## 3.1 Mathematical formulation

The following section presents a detailed process and mathematical formulation for developing a transit control system using DRL technology, building upon the assumptions and perspectives established earlier.

To facilitate the development of a DRL-based control system, a realistic DRL environment is established, as illustrated in Figure 2, incorporating both historical and real-time data. This environment comprises four interactive modules: the scheduled bus motion module, space-time varying disturbance module, actual bus motion module, and error dynamic module. The scheduled bus motion module outlines the planned bus operation according to the schedule, denoted as the scheduled arrival time at the station

(Equation 1). However, bus operations can be disrupted by spatiotemporal changes. To mitigate this issue and accommodate delays and passenger demand uncertainty, the space-time varying disturbance module is formulated, more effectively utilizing both bus history and real-time traffic information in a parametric or non-parametric manner. The actual bus motion module is subsequently developed to characterize the real bus operation, specifically represented as the actual arrival time at the station (Equation 3). Lastly, the error dynamic module is assembled to concentrate on the discrepancy between actual bus motion and scheduled motion. Relevant notations in the proposed approach are defined in Table 1.

Table 1. Notations in the proposed approach.

| Symbol | Definition |
|---|---|
| $i$ | Indexes to denote bus, $i = 1,2 \ldots, M$ |
| $j$ | Indexes to denote positions, $j = 1,2 \ldots, N$ |
| $k$ | Indexes to denote signals, $k = 1,2, \ldots N_j$ |
| $t_j^i$ | The planned arrival time for bus $i$ at position $j$ |
| $H$ | The planned headway |
| $\beta_j$ | The expected passenger demand rate at position $j$ |
| $r_j$ | The average travel time from position $j$ to position $j+1$ |
| $s_j$ | The slack time planned in the schedule |
| $w_j^i$ | The disturbance delay to the travel time of bus $i$ from position $j$ position $j+1$ |
| $\Delta \beta_j^i$ | The uncertainty to the passenger demand rate for bus $i$ at position $j$ |
| $u_j^i$ | The control force for bus $i$ at position $j$ |
| $u_{j(b),b}^i$ | The control force using bus holding at bus station |
| $u_{j(k),k}^i$ | The control force using transit signal adjustment |
| $u_{j(c),c}^i$ | The control force using inter-station curies speed control |
| $\alpha_{3,i,b}$ | The coefficient for control force using bus holding |
| $\alpha_{3,i,k}$ | The coefficient for control force using transit signal priority |
| $\alpha_{3,i,c}$ | The coefficient for control force using cruise speed control |
| $V_{j,k,n}/C_{j,k,n}$ | The V/C ratio of the traffic phase for the major street movement $n$ at intersection $k$ of the station $j$ |
| $a_j^i$ | The actual arrival time of bus $i$ at position $j$ |
| $h_j^i$ | The actual headway of bus $i$ at position $j$ |
| $\tilde{\beta}_j^i$ | The actual passenger demand rate for bus $i$ at position $j$ |
| $e_j^i$ | The schedule deviation of bus $i$ at position $j$ |
| $d_j^i$ | The headway deviation of bus $i$ at position $j$ |

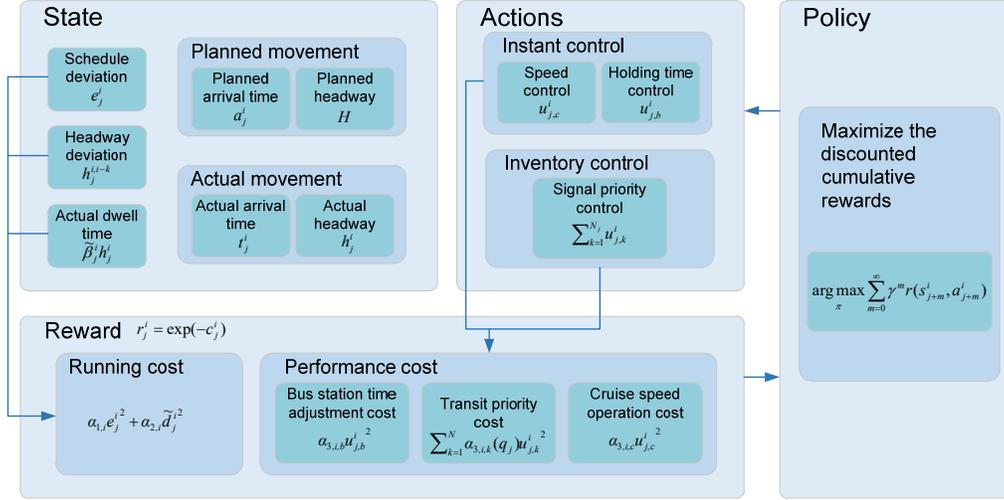

Figure 2. The DRL method design of the paper.

Firstly, the scheduled arrival time for bus $i$ at position $j + 1$ is defined as the sum of the previous arrival time at the last station, average arrival time, dwell time, and slack time. The scheduled bus motion module can be expressed as follows:

$$t_{j+1}^i = t_j^i + \beta_j H + r_j + s_j \tag{1}$$

The expected dwell time for the bus to serve passengers at position $j$ is defined as $\beta_j H$. $H$ is the planned headway and $\beta_j$ is the expected demand rate. The planned headway is defined as $H = t_j^i - t_j^{i-1}$. $r_j$ is the average travel time.

In the space-time varying disturbance module, the disturbance sources of bus movement we consider are mainly divided into two parts, including the delay disturbance $w_j^i$ and the passenger demand uncertainty $\Delta \beta_j^i$, which can be built as mathematical or empirical distributions based on historical statistics. The delay disturbance could result from multiple reasons, such as traffic conditions, road conditions, accidents, environment, and traffic signals. The actual demand rate is assumed as time-invariant and irrelative to locations, which is expressed as:

$$\tilde{\beta}_j^i = \beta_j + \Delta \beta_j^i \tag{2}$$

Based on the scheduled bus motion module, the disturbance module, and the control force $u_j^i$, the actual bus motion module can be written as:

$$a_{j+1}^i = a_j^i + \tilde{\beta}_j^i h_j^i + r_j + u_j^i + w_j^i \tag{3}$$

Similarly, the actual headway writes $h_j^i = a_j^i - a_j^{i-1}$. $u_j^i$ is the general control force that includes multiple strategies, such as inter-station operation speed control, early departure, and transit signal priority adjustment. It writes:

$$u_j^i = u_{j(b),b}^i + u_{j(k),k}^i + u_{j(c),c}^i \tag{4}$$

The control force $u_{j(b),b}^i$ represents the time adjustment for bus holding ($u_{j(b),b}^i > 0$) at a bus station. The control force $u_{j(k),k}^i$ represents the time adjustment for bus $i$ from position $j$ to position $j+1$ benefit from transit signal adjustment at signalized intersections. The control force $u_{j(c),c}^i$ represents the time adjustment for bus $i$ via an inter-station operation speed control from position $j$ to position $j+1$. Since controlling speed may influence other traffic on the same road, the difficulty of controlling speed may be larger than that at the bus station or a signalized intersection. the discount factor is set larger than others.

The error dynamic module shows the deviations between the actual bus motion and the scheduled bus motion. Two types of errors are defined; they are schedule deviation and headway deviation.

The schedule deviation $e_j^i$ is defined as the difference of scheduled arrival time and actual arrival time:

$$e_j^i = a_j^i - t_j^i \tag{5}$$

The headway deviation is defined as the difference between planned headway and actual headway:

$$d_j^i = h_j^i - H \tag{6}$$

Information from downstream buses is considered in the control framework to better optimize the buses' motions. The headway deviation between bus $i$ and bus $i - k$ is defined as:

$$d_j^{i,i-k} = h_j^{i,i-k} - kH \tag{7}$$

$$h_j^{i,i-k} = a_j^i - a_j^{i-k} \tag{8}$$

A distributed DRL-based control strategy is formulated, drawing upon the data-driven bus transit system. By determining the control force for each bus at every station, signalized intersection, and inter-station area, both schedule adherence and headway regularity are effectively maintained.

*3.2 DRL scheme*

The reinforcement learning model is founded on a Markov Decision Process (MDP). The model comprises two interactive entities: the environment (simulation platform) and the RL agent (bus control algorithm). As presented in Figure 2, these entities encompass state, action, policy, and reward, denoted as (*s, a, π, r*).

A physics-informed DRL-based distributed control framework (Shi et al., 2022) is designed for cooperation within a multi-agent bus network. The data collected by the subject bus includes its own state, the state of the station ahead, traffic conditions, and signal timing information along the road. The state of $k$ downstream buses adopt the control force $u_j^i$ output by the DRL-based model $k$ ($M_k$) at each station, signalized intersection, and inter-station area. During the first round of a day, the number of downstream buses may be less than $k$. In such cases, the subject bus gathers information from all available downstream buses. If no downstream bus is present, a dummy bus without deviation from the schedule is designated as the downstream bus.

The physical fused DRL state ensures buses remain in proximity to the pre-defined equilibrium state of bus operations (i.e., the scheduled bus state), while simultaneously achieving a consensus performance among bus platoon. In the context of bus operation control, the equilibrium point serves as the ideal state of the bus while in motion, which guides the exploration direction in DRL training to help develop a robust control policy. The control objectives are guided by the reward to be optimized and the reward function is specifically designed taking into account the physics-informed DRL state. To ensure that buses remain close to equilibrium, the DRL agent must engage in action exploration, policy updates, and convergence. These DRL components are explained below.

*3.2.1 State definition*

The definition of the state is critical because it allows the agent to plan the next control based on the current bus operation and control objectives. The DRL state for the bus control task is physics-informed and integrates multiple sources of information from downstream buses and road traffic environment based on the equilibrium concept. As shown in **Error! Reference source not found.**, the state $s$ incorporates three elements: the schedule deviation $e_j^i$, the weighted headway deviation $\tilde{d}_j^i$ that integrates downstream bus information, and the actual dwell time $\tilde{\beta}_j^i h_j^i$ related to the actual demand rate. Thus, for a given bus $i$ at a location $j$, $j$ shows a specific location whether it is at a bus station, inner-intersection, or at a signalized intersection. In the model, the length of the block between two positions $j$ and $j+1$ is unique. the state is defined as $s_j^i = [e_j^i, \tilde{d}_j^i, \tilde{\beta}_j^i h_j^i]$, where $\tilde{d}_j^i$ is defined as the weighted average of the deviation of heads that leads to a consensus and stable transit system (an equilibrium headway):

$$\tilde{d}_j^i = w_{i-1} d_j^i + w_{i-2} d_j^{i,i-2} + \cdots + w_{i-k} d_j^{i,i-k} \tag{9}$$

The weighted coefficient $w_{i-m}$ shows that the closer downstream bus has a larger impact on the controlled bus than the further ones, which is defined as below:

$$w_{i-m} = \begin{cases} 1/2^m, & 1 \leq m \leq k-1 \\ 1/2^{m-1}, & m = k \end{cases} \tag{10}$$

### 3.2.2 Action definition

For bus $i$ at position $j$, when the RL agent receives the state information $s_j^i$, it outputs actions $a$ (which corresponds to $u_j^i$) to control the bus according to the current policy $\pi$. The action is specifically designed as a combined application of bus holding, bus operating speed adjustment, and traffic signal priority, tailored to adjust bus operations in accordance with each implementation location (i.e., bus station, inner-station roads, and signalized intersections).

### 3.2.3 Reward definition

The reward $r$ is employed to represent the control objectives. Three targets – schedule deviation, weighted headway deviation, and control force – are minimized to maintain schedule adherence and headway regularity under fluctuating disturbances, while utilizing a low-cost control force.

Specifically, the immediate reward $r_j^i$ for bus $i$ at position $j$ is designed concerning three types of costs. An exponential function is to limit the value in the boundary $[0, 1]$, given by:

$$r_j^i = exp(-c_j^i) \tag{11}$$

The running cost of the three objectives are represented by quadratic forms to improve the training performance:

$$c_j^i = (x_j^i)^T Q_i x_j^i \tag{12}$$

$$c_j^i = \alpha_{1,i} {e_j^i}^2 + \alpha_{2,i} {\tilde{d}_j^i}^2 + \alpha_{3,i,b} {u_{j(b),b}^i}^2 + \alpha_{3,i,k}(q_j) {u_{j(k),k}^i}^2 + \alpha_{3,i,c} {u_{j(c),c}^i}^2 \tag{13}$$

where $(x_j^i)^T = [e_j^i, \tilde{d}_j^i, u_j^i]$; $Q_i$ is a positive definite diagonal coefficients matrix, designed as the diagonal matrix below:

$$Q_i = \begin{bmatrix} \alpha_{1,i} & & & & \\ & \alpha_{2,i} & & & \\ & & \alpha_{3,i,b} & & \\ & & & \alpha_{3,i,k}(q_j) & \\ & & & & \alpha_{3,i,c} \end{bmatrix}, \alpha_{1,i}, \alpha_{2,i}, \alpha_{3,i}, \alpha_{3,i,k}, \alpha_{3,i,c} > 0 \tag{14}$$

where $\alpha_{1,i}$ represents the coefficient for schedule deviation; $\alpha_{2,i}$ represents the coefficient for headway deviation, and $\alpha_{3,i,b}$, $\alpha_{3,i,k}$, $\alpha_{3,i,c}$ represent the control force's coefficient. Large values of $\alpha_{1,i}$ and $\alpha_{2,i}$ may enhance the schedule adherence and the headway regularity, but could overlook the control force cost and the implementation difficulty of each control strategy. Conversely, large values of $\alpha_{3,i,b}$, $\alpha_{3,i,k}$ and $\alpha_{3,i,c}$ may lead to suboptimal control performance. Consequently, it is crucial to carefully determine the values of these coefficients by taking various factors into account.

It is important to note that the implementation of transit signal priority must be integrated with the traffic flow conditions at the intersection, while striving to ensure that the intersection's traffic capacity does not decrease or only experiences a slight decline. In other words, since a larger flow can result in reduced flexibility when adjusting the signal, we introduce an additional coefficient term $q_j$ for the signal adjustment control force, representing the volume cost at position $j$. The volume cost is defined as follows:

$$q_j = \sum_k c_{j,k} \tag{15}$$

$$c_{j,k} = \frac{\sum_{n=1}^{N} V_{j,k,n}/C_{j,k,n}}{V_{j,k,m}/C_{j,k,m}} \tag{16}$$

where $c_{j,k}$ represents the volume cost of the intersection $k$ at station $j$; $V_{j,k,m}/C_{j,k,m}$ represents the V/C ratio of the traffic phase for the major street movement $m$; $\sum_{n=1}^{N} V_{j,k,n}/C_{j,k,n}$ represent the summation of the V/C ratio for traffic phases on all sides. Consequently, it is incorporated into the model as an influential variable that affects the feasibility of implementing signal priority.

The three control forces are designed to distinguish between multiple time adjustment strategies at three distinct locations. The control range for each type of control has been constructed by carefully considering the feasibility of bus station time adjustment, bus priority time adjustment, and inner-interstation time adjustment.

For the location of bus station, the control force by using time adjustment at bus stations is

$$u_{j(b),b}^i \in [0, t_{j(b),max}] \tag{17}$$

$t_{j(b),max}$ is the maximal extra time a bus can park at a station. The early leave strategy is disabled due to the implementation difficulties considering meeting the demand at the bus station. $t_{j(b),max}$ is set at 20 seconds in this paper.

For the signal intersection position, the control force using transit signal adjustment is

$$u^i_{j(k),k} \in [-t_{j(k),max}, t_{j(k),max}] \qquad (18)$$

where $t_{k,i,max}$ is the absolute value of maximal time saving/extending a bus priority strategy at signalized intersections that can be provided for the subject bus at position $j$. $t_{k,i,max}$ is set at 20 seconds in this paper.

The cruise speed operating control force for adjusting the inner-interstation operation time is

$$u^i_{j(c),c} \in [-t_{j,min}, t_{j,max}] \qquad (19)$$

$t_{j(c),min}$ is the absolute value of min time saving a bus can get by increasing speed at the position $j$ to the next position, and $t_{j(c),max}$ is the absolute value of max time relax a bus can get via decreasing the speed at the position $j(c)$. $t_{j,min}$ and $t_{j,max}$ are determined by the following equations:

$$t_{j,min} = M_{j,j+1}/v_{j,min} - M_{j,j+1}/r_j \qquad (20)$$

$$t_{j,max} = M_{j,j+1}/r_j - M_{j,j+1}/v_{j,max} \qquad (21)$$

where $M_{j,j+1}$ represents the distance between the position $j$ and position $j + 1$; $v_{j,min}$ denotes the minimum average speed that can be decreased; $v_{j,max}$ represents the maximum speed that can be reached.

*3.2.4 Policy definition*

The policy $\pi$ is an implicit function that is updated through the training process to achieve optimal control. An infinite-horizon optimal control problem is formulated based on the reward function. For bus $i$ at station $j$, the optimal policy $\pi^*$ aims to maximize the discounted cumulative rewards over the infinite time horizon:

$$\pi^* = \arg\max_{\pi} \sum_{m=0}^{\infty} \Upsilon^m \, r\big(s_{j+m}^i, a_{j+m}^i\big) \qquad (22)$$

where $r\big(s_j^i, a_j^i\big)$ represents the reward function.

*3.3 DRL training*

To solve the optimization problem defined in Equation 22, this paper employs a DPPO to enhance training performance. The DPPO algorithm is proposed based on the proximal policy optimization (PPO) algorithm (Heess et al., 2017), while emulating the network structure of the asynchronous advantage actor-critic (A3C) algorithm (Mnih et al., 2016).

      The PPO algorithm aims to address the issue of the poorly determined learning rate in the PG algorithm. If the learning rate is too large, the learned strategy may struggle to converge. On the other hand, if the learning rate is too small, the process will be time-consuming. The PPO algorithm employs the ratio of the new strategy $\pi_\theta$ and the old strategy $\pi_{\theta_{old}}$ to limit the update range of the new strategy, making the algorithm less sensitive to a slightly larger learning rate, thereby improving the algorithm's efficiency.

      In DPPO, a global network and a multi-agent network are present, with each agent interacting with its own independent environment to gather data. The global network updates parameters based on a batch of data collected by all agents, and the agents continue to interact with the environment using the latest strategy. Overall, the DPPO

algorithm exhibits superiority in terms of sample efficiency, performance, and algorithm convergence.

The control model is trained using the DPPO algorithm, allowing the policy to be updated in order to achieve the desired control performance. Specifically, the DPPO agent interacts with the environment to control six buses in real-time. Each parallel agent receives the state $s_j^i$ of the bus at station $j$, after which the global actor network updates the policy and outputs the control force $u_j^i$ to regulate the bus's operation to the next station. The reward is obtained by calculating the reward function, and subsequently, the bus's state is updated.

To train the model effectively, the hyperparameters of the DPPO algorithm are configured as follows: The minibatch is set to T=256, the clipping value is set to $\varepsilon = 0.2$, and the discount factor is $Y = 0.99$. Furthermore, the learning rate for both the actor network and critic network is set to 1e-5.

Once sufficient data is collected, the actor network and critic network are updated separately to optimize the policy and minimize the critic loss. The objective function $L^{CLIP}(\theta)$ is maximized to update θ, as demonstrated in Equation (23), which subsequently updates the actor network responsible for action selection.

$$L^{CLIP}(\theta) = \hat{E}_t[min\left(p_t(\theta)\hat{A}_t, clip(p_t(\theta), 1 - \varepsilon, 1 + \varepsilon)\right)\hat{A}_t] \quad (23)$$

where $p_t(\theta) = \pi_\theta(a_t|s_t)/\pi_{\theta_{old}}(a_t|s_t)$ represents the probability ratio between the update strategy $\pi_\theta$ and the old strategy $\pi_{\theta_{old}}$. To improve convergence, $p_t(\theta)$ is constrained to be within $1 - \varepsilon$ to $1 + \varepsilon$ by the $clip(p_t(\theta), 1 - \varepsilon, 1 + \varepsilon)$ function. The advantage function $\hat{A}_t$ is expressed as follows:

$$\hat{A}_t = R_t - V_\phi(s_i^t) \quad (24)$$

where $R_t$ is the discounted cumulative reward for $T$ time steps:

$$R_t = \sum_{m=0}^{T-1} \gamma^m r_i^{t+m} + \gamma^T V_\phi(s_i^{t+T}) \tag{25}$$

The critical network is updated by minimizing the critic loss $L_c(\phi)$ to better evaluate the output action:

$$L_c(\phi) = \hat{E}_t \left( R_t - V_\phi(s_i^t) \right)^2 \tag{26}$$

During training, a single time step update signifies a bus traveling from one stop to the next, with the number of training episodes set to 2000 for $M_1 \sim M_5$. This paper records the rewards of the proposed model in units of episodes, where each training episode encompasses the process of six buses completing a full loop of 20 stations. Consequently, the number of time steps for each episode, $N$, is set at 20. The reward trajectories in Figure 3 display the reward convergence for $M_1$ to $M_5$, respectively. As illustrated in the figure, the reward gradually converges after training for 700 episodes under various downstream bus number scenarios, demonstrating excellent convergence in the training process.

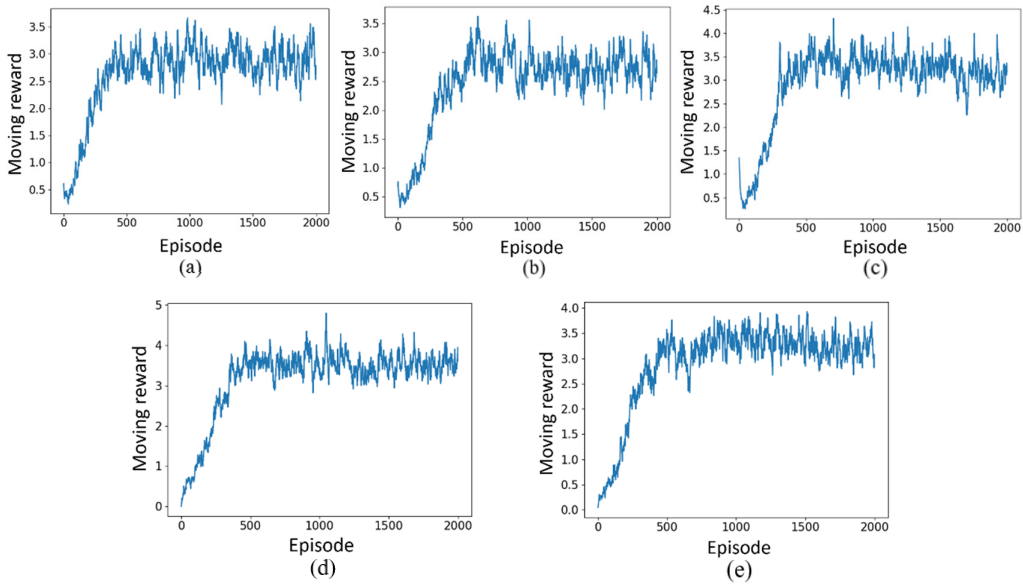

Figure 3. Reward trajectories of the proposed models, the reward trajectory of $M_1$ (a), the reward trajectory of $M_2$ (b), the reward trajectory of $M_3$ (c), the reward trajectory of $M_4$ (d) and the reward trajectory of $M_5$ (e).

## 4. Numerical experiments

### 4.1 Experiment setup

To showcase the effectiveness of the proposed DRL-based bus control strategy, we carried out numerical experiments using bus operation data from Beijing Bus Line 16 in this section. The DRL-based bus controller was developed using TensorFlow.

In the experiment, the bus system is designed as a looped structure with 20 stations, with 19 buses operating in the system. After reaching the terminal, buses continue to loop and start a new trip. The number of signalized intersections $k$ is set to 1, and traffic conditions are varied by setting different traffic volumes, while the number of downstream buses n is set to 5. For the disturbance of bus movement, we set the delay disturbance $w_j^i$ to follow a truncated normal distribution within the boundary of [-5s, 30s] to simulate traffic condition fluctuations, and set the passenger demand uncertainty $\Delta\beta_j^i$ to follow a uniform distribution, represented by the interval [-0.02, 0.02].

After analyzing the historical driving data of Beijing Line 16, we assume a planned headway of H = 300s. Based on the actual bus station situation and in accordance with settings from existing literature (Daganzo, 2009; Li et al., 2019), the passenger demand rate $\beta_j$ for each station is set within the range of [0.03, 0.12] to reflect realistic passenger demand fluctuations. We assume that the average travel time $r_j$ between two adjacent stations ranges between 240s and 260s. Table 2 below presents the details of the average travel time for each pair of adjacent stations and the average passenger demand rate for each station. The planned slack time $s_j$ for station $j$ is set at 10s. For the positive definite diagonal coefficient matrix $Q_i$, the coefficient for schedule deviation $\alpha_{1,i}$ and the coefficient for headway deviation $\alpha_{2,i}$ are both 0.01. The control force coefficients $(\alpha_{3,i,b}, \alpha_{3,i,k}, \alpha_{3,i,c})$ are all set to 0.01.

Table 2. Profiles of average travel time and average passenger demand rate.

| Station | Average travel time (s) | Average passenger demand rate | Station | Average travel time (s) | Average passenger demand rate |
|---|---|---|---|---|---|
| 1 | 257.0 | 0.08 | 11 | 246.0 | 0.05 |
| 2 | 253.0 | 0.05 | 12 | 240.0 | 0.11 |
| 3 | 257.0 | 0.03 | 13 | 256.0 | 0.04 |
| 4 | 259.0 | 0.09 | 14 | 250.0 | 0.12 |
| 5 | 246.0 | 0.07 | 15 | 242.0 | 0.07 |
| 6 | 247.0 | 0.11 | 16 | 251.0 | 0.08 |
| 7 | 260.0 | 0.06 | 17 | 242.0 | 0.05 |
| 8 | 256.0 | 0.11 | 18 | 250.0 | 0.11 |
| 9 | 240.0 | 0.05 | 19 | 248.0 | 0.08 |
| 10 | 252.0 | 0.10 | 20 | 257.0 | 0.04 |

After setting the parameters, experiments are conducted in the following three aspects to evaluate the effectiveness of the proposed bus control strategy:

(1) General performance comparison;

(2) Bus control under different traffic volume conditions;

(3) Sensitivity analysis (Single control force experiment).

In the experiment, buses operate on a circular route and complete two loops. After reaching the terminal, they continue the loop to start a new trip. The station index increases continuously (from 1 to 40) as the bus completes a lap.

*4.2 General performance comparison*

In this section, we conduct a simulation experiment to evaluate the performance of our proposed DRL-based multi-strategy bus control method in comparison to traditional bus control methods. The experiment considers a general case and aims to demonstrate the similarity in performance to the method proposed by Shi et al. (2022). Specifically, we compare our approach against the cases of no control, schedule-based control strategy, and headway-based control strategy. Our analysis of the experimental results highlights the effectiveness of the proposed method in aligning the actual bus operation with the planned operation across various scenarios.

*4.2.1 No control situation*

The experimental results for the no-control scenario are depicted in Figure 4. Figure 4a portrays the bus trajectory, where yellow dashed lines represent the scheduled bus trajectories, and dark blue solid lines signify the actual bus trajectories. The red circle emphasizes the intersection of the solid lines, indicating instances of bus bunching. The findings reveal that multiple factors, including challenging environmental conditions and driving behavior, contribute to a substantial divergence between the actual and scheduled bus trajectories over time. This results in frequent bus bunching in the absence of control measures.

For a clearer representation of the schedule and headway deviations, Figure 4b and Figure 4c are presented. These illustrations show that deviations escalate from one station to the next, primarily due to recurrent delay disturbances and considerable passenger demand uncertainty, reaching peak deviations of more than 200 seconds. The deviations reset to zero upon arriving at the terminal station.

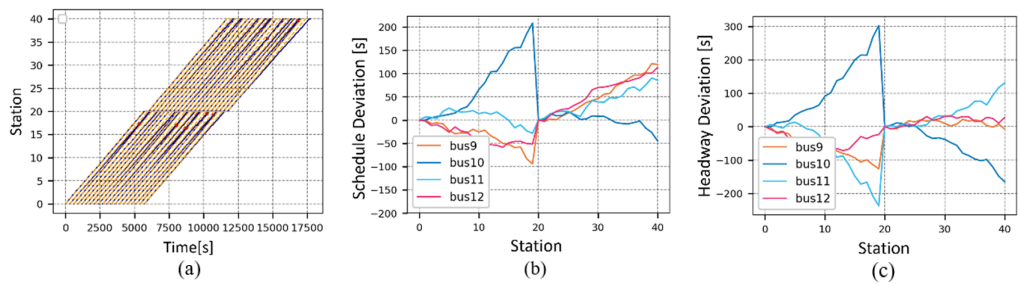

Figure 4. Bus trajectories (a), schedule deviation (b), and headway deviation (c) of no control.

*4.2.2 Schedule-based method*

Figure 5 depicts the results obtained through the implementation of the schedule-based method. As shown in Figure 5a, the frequency of bus bunching occurrences has significantly decreased. Moreover, Figures 5b and 5c illustrate the deviations, which were moderately reduced to less than 130 s.

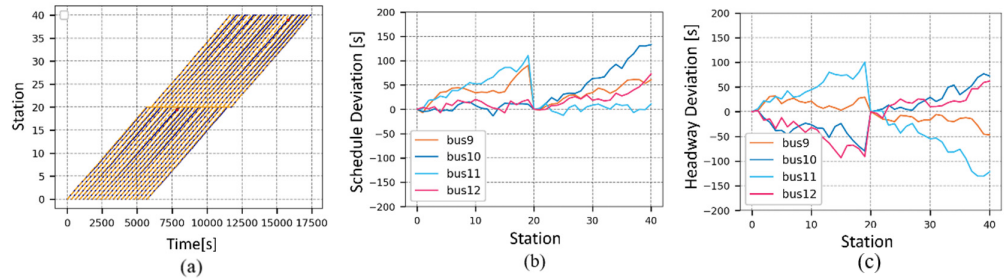

Figure 5. Bus trajectories (a), schedule deviation (b), and headway deviation (c) of schedule-based method.

*4.2.3 Headway-based method*

The bus control results obtained through the headway-based method are presented in Figure 6. As illustrated in Figure 6a, the solid blue line representing the actual bus trajectories does not intersect, indicating that the headway-based method effectively maintains the minimum allowable headway to prevent bus bunching. Additionally, Figures 6b and 6c demonstrate a further reduction in deviations to less than 120 s.

While the existing widely used bus control strategies have proven effective in reducing the frequency of bus bunching and improving bus operation efficiency, they solely rely on a single approach to control the bus and fail to account for real-time environmental information of the bus system, thereby highlighting their limitations.

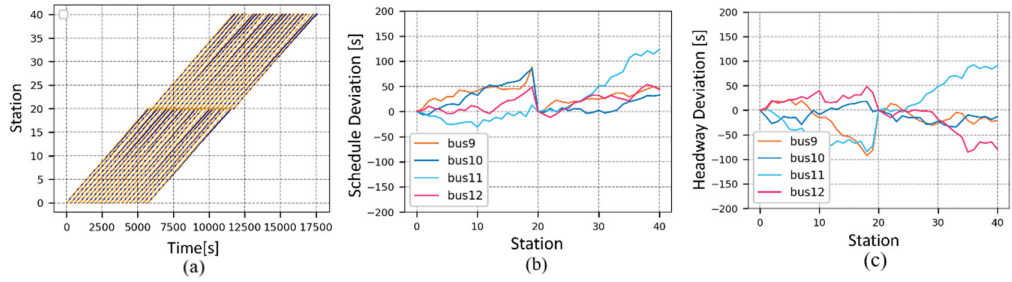

Figure 6. Bus trajectories (a), schedule deviation (b), and headway deviation (c) of headway-based method.

*4.2.4 Proposed method*

Figure 7 provides a comprehensive representation of the experimental results obtained using the proposed method. The bus trajectories depicted in Figure 7a exhibit a high degree of conformity between the scheduled trajectory and the actual trajectory. The schedule deviation trajectories (Figure 7b) and headway deviation trajectories (Figure 7c) also indicate that deviations can be maintained within 35 seconds, which is considerably lower than those observed in other scenarios.

      The proposed method combines bus operation history with real-time traffic information, taking into account the effects of delay disturbances and passenger demand uncertainty. Furthermore, it integrates and applies various control strategies and sets three control variables, including dwell time at a stop, speed during inter-station travel, and signal priority. These variables have shown exceptional performance in preserving the accuracy of the scheduled time and the regularity of headways. In comparison to other scenarios, the proposed method demonstrates superior performance.

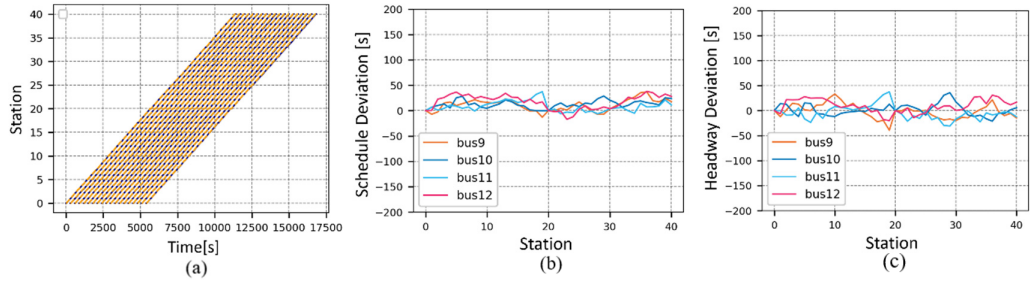

Figure 7. Bus trajectories (a), schedule deviation (b), and headway deviation (c) of the proposed method.

## 4.3 Bus control with different traffic volume conditions

This scenario examines the performance of the method under various congestion levels by assuming different traffic volumes on the main street. To simulate diverse traffic conditions, we assigned distinct traffic costs to specific stations. In particular, we set the traffic cost to be very high for some stations to demonstrate the interaction between the three control strategies under changing traffic volume conditions and to understand the control force patterns of these strategies. Table 3 provides the volume cost for each station, and the corresponding line chart is presented in Figure 8.

Table 3. The setting of traffic cost for each station.

| Station | 1 | 2 | 3 | 4 | 5 | 6 | 7 | 8 | 9 | 10 |
|---|---|---|---|---|---|---|---|---|---|---|
| Volume cost | 1 | 4 | 10 | 3 | 80 | 20 | 7 | 20 | 10 | 3 |
| Station | 11 | 12 | 13 | 14 | 15 | 16 | 17 | 18 | 19 | 20 |
| Volume cost | 1 | 4 | 20 | 3 | 30 | 20 | 80 | 6 | 10 | 3 |

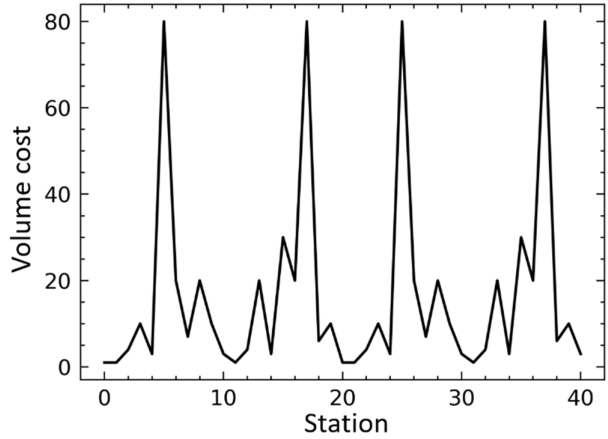

Figure 8. Volume cost for each station.

Figure 9 displays the control forces of the three control strategies for Bus 9. When the volume cost is low, the bus holding control force tends to be larger. This occurs because the bus often arrives at the station ahead of the scheduled time, leading to the implementation of the bus holding strategy. Conversely, when the volume cost is high, this strategy is not employed, resulting in a control force of 0.

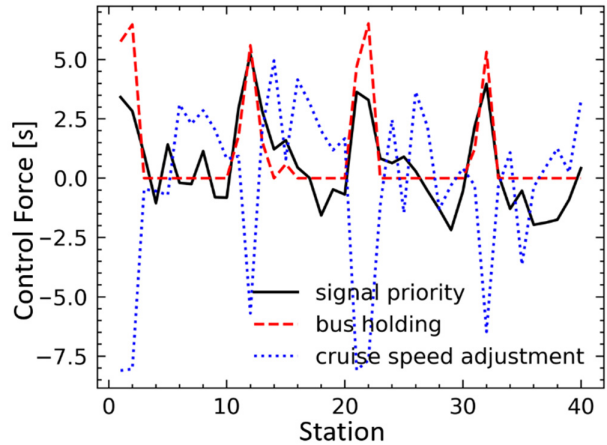

Figure 1 The control force of the three control strategies.

In terms of cruising speed, when the traffic cost is low, the adjustment magnitude for cruising speed is predominantly negative, reaching a minimum value of -8s. This suggests that the bus has considerable flexibility to accelerate in order to better adhere to the planned schedule. Conversely, under high traffic cost conditions, the bus is primarily adjusted to decelerate.

The control force for signal priority exhibits a trend opposite to that of the volume cost. Specifically, when the traffic volume cost is low, the maximum signal priority control force can surpass 5s. However, when the volume cost reaches its peak value, the control force oscillates around 0, with a minimum of -2s.

Figure 10 displays minimal fluctuations in schedule deviation and headway deviation during bus operation, with deviations maintainable within 35s. These results indicate that the proposed bus control method is highly effective and robust when faced with varying traffic volume conditions. Consequently, the method showcases outstanding performance in bus control.

Moreover, the results indicate that the effectiveness of the three control strategies varies with changing traffic conditions. However, the three strategies demonstrate an exceptional ability to complement each other, leading to enhanced control performance. These findings offer empirical evidence supporting the proposed strategy's resilience to fluctuations in traffic volume conditions.

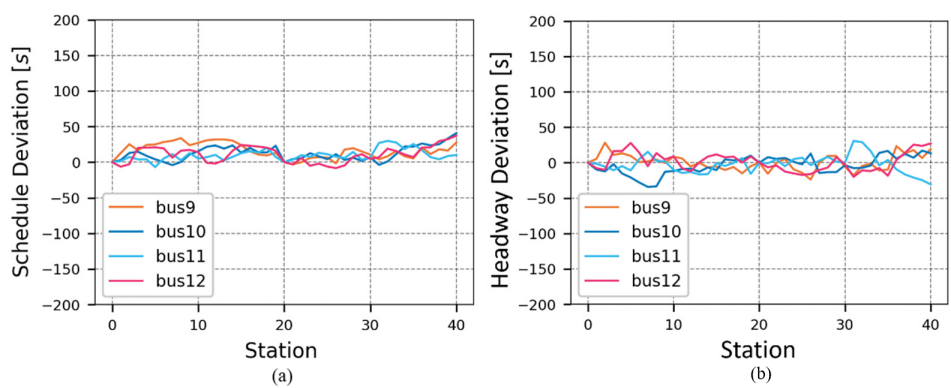

Figure 10. The schedule deviation (a) and headway deviation (b) trajectories under different traffic volume conditions.

## 4.4 Sensitivity analysis (Single control force experiment)

In this section, we explore scenarios where only one of the three control forces is employed. Sensitivity analyses are conducted for each case, evaluating the degree to which the control needs to be applied to achieve optimal performance under various levels of disturbance. It is important to note, however, that relying solely on a single control force might not yield satisfactory results due to factors such as high volume cost and constrained control forces. Experimental results for the implementation of individual bus holding control, signal priority, and cruise speed control under high traffic conditions are presented in Figure 10, Figure 11, and Figure 12, respectively. The traffic volume cost settings used in the experiments are detailed in Table 4.

Table 4. The setting of traffic cost for each station.

| Station | 1 | 2 | 3 | 4 | 5 | 6 | 7 | 8 | 9 | 10 |
|---|---|---|---|---|---|---|---|---|---|---|
| Volume cost | 80 | 60 | 90 | 70 | 80 | 90 | 80 | 90 | 60 | 80 |
| Station | 11 | 12 | 13 | 14 | 15 | 16 | 17 | 18 | 19 | 20 |
| Volume cost | 50 | 90 | 100 | 30 | 90 | 80 | 70 | 90 | 80 | 70 |

### 4.4.1 Bus holding control only with high volumes

To showcase the effectiveness of bus holding control, we first analyze the schedule deviation (Figure 11a) and headway deviation (Figure 11b) in scenarios where only this control force is employed. These figures indicate that deviations can reach up to 140 seconds and accumulate downstream along the route. Furthermore, Figure 11c displays the bus holding control force trajectory for bus 9, with the control force boundary defined as [1.4 s, 6.5 s].

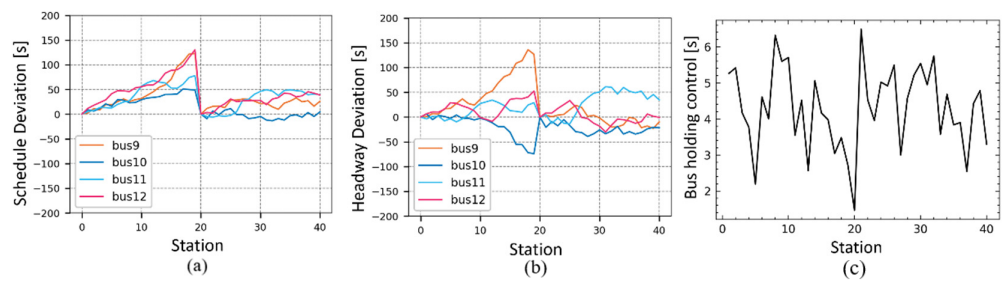

Figure 11. The schedule deviation (a), headway deviation (b) and bus holding control force (c) trajectories under bus holding control only with high volumes.

*4.4.2 Cruise speed control only with high volumes*

When utilizing solely cruise speed control, the schedule and headway deviations remain under 90 s and 100 s, respectively, as shown in Figure 12a and 12b. The force of cruise speed control for bus 9 ranges from -3.4 to 1.0 s, indicating limited room for adjusting cruise speed during high volume periods.

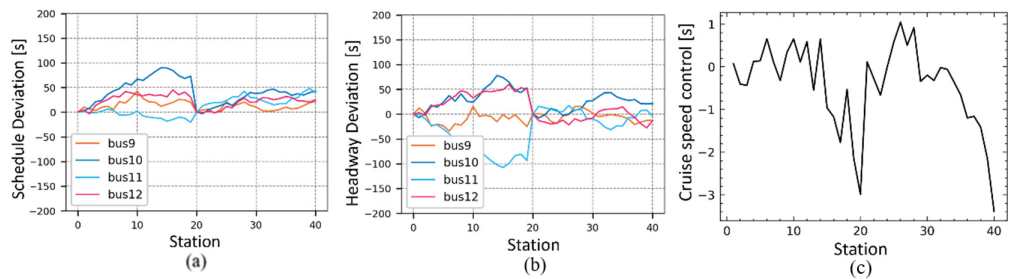

Figure 12. The schedule deviation (a), headway deviation (b) and bus holding control force (c) trajectories under cruise speed control only with high volumes.

*4.4.3 Signal priority only with high volumes*

The results presented in Figures 13a and 13b show that signal priority can decrease both schedule deviation and headway deviation to approximately 50 seconds. Notably, the signal priority control force for bus 9 exhibits a relatively broad range, with values ranging from -2.8 s to 6.9 s.

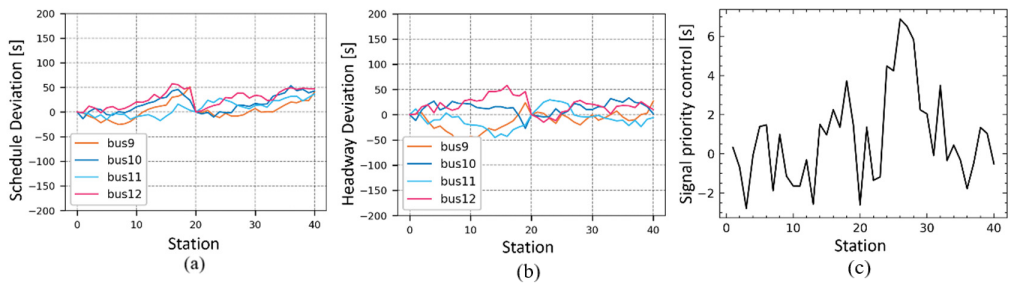

Figure 13. The schedule deviation (a), headway deviation (b) and bus holding control force (c) trajectories under signal priority only with high volumes.

*4.4.4 Proposed control with high volumes*

Finally, the control effect of the proposed bus control method based on multi-strategy fusion is demonstrated. As shown in Figure 13a and 13b, the deviations in the proposed system are greatly reduced compared to the first three cases, with deviations being controlled within 45s.

In conclusion, it can be inferred that employing any of the aforementioned three strategies singularly has a beneficial impact on reducing both schedule deviation and headway deviation, and each strategy provides some level of alleviation to the bus bunching problem.

When faced with high traffic costs, the single control strategy is insufficient to deliver optimal performance, particularly when bus holding control is used in isolation. However, the proposed control, which incorporates three different control strategies, operates synergistically to yield superior performance while ensuring the control's flexibility. Additionally, the control forces range of bus 9, which spans from -3 to 6 s across the four cases, indicates that the strategies apply to vehicles that arrive earlier or later than the schedule. Moreover, the control actions are straightforward to implement since the control forces are kept within a short range.

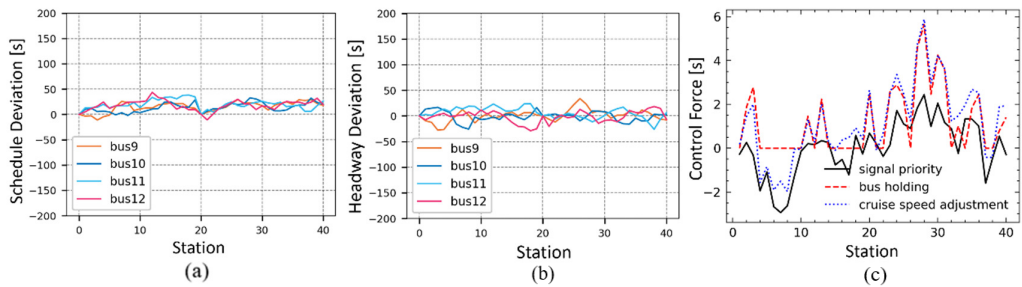

Figure 14. The schedule deviation (a), headway deviation (b) and bus holding control force (c) trajectories under proposed control with high volumes.

## 5. Conclusion

Accurate, real-time, and robust bus operation control is an important support for alleviating bus bunching. This paper proposes a multi-strategy bus control strategy based on distributed DRL, which comprehensively considers the accuracy of bus arrival time, the regularity of headway, and the consensus in the multi-agent system, and captures the uncertainty in the process of bus operation. Specially, we consider the influence of traffic volume and signals and set three factors as control variables, which are the dwell time at a stop, speed during inter-station, and priority at signals. Through the integrated application of various control strategies, the frequency of bus bunching can be significantly reduced and the efficiency of bus operation can be improved.

The construction of a realistic DRL environment is based on historical data and real-time traffic information obtained by CAV technology, and a distributed control framework is designed on this basis. Each bus utilizes the fusion state information of downstream buses to achieve the multi-agent consensus and maintain the stability of the bus system. Training through the DPPO algorithm to update the control strategy. The DRL agent can achieve the optimal control force through the application of multiple strategies fusion to achieve the desired control performance. At the same time, we consider the feasibility of implementing the strategies to ensure that it is suitable for different traffic conditions.

A series of numerical experiments are conducted to comprehensively evaluate several aspects of the proposed control method. Firstly, the control performance is compared with the existing mainstream bus control methods. The results show that our proposed method excels in maintaining the accuracy of scheduling time and the regularity of headway. Besides, we investigate the performance of the method under different congestion levels and the results show that the three control forces cooperate well with each other, which proves that the proposed strategy is hardly affected by the fluctuation of traffic conditions. The last experiment shows that a single control strategy is limited and limited control can not achieve good results in the case of high traffic costs. In contrast, the proposed method is more flexible in control, and the combination of the three strategies shows better performance.

Several research topics could be carried out in the future. First, it is necessary to establish a bus control model that can handle multiple bus routes. Second, a more elegant passenger demand model considering the number of alighting passengers and boarding passengers as well as the limitation of bus capacity could be developed. Third, some deep learning algorithms (e.g., Zeng et al., 2022) could be adopted to build an accurate and reliable bus arrival time forecasting model in order to further develop a proactive bus control strategy.

## Author Contributions

**Qinghui Nie:** Conceptualization, Methodology, Experiment, Analysis, Writing-Original draft preparation. **Jishun Ou**: Conceptualization, Methodology, Writing-Review and Editing, Revision. **Haiyang Zhang:** Methodology, Data curation, Writing-Review and Editing. **Jiawei Lu:** Writing-Review and Editing, Revision. **Shen Li:** Writing-Review and Editing, Revision. **Haotian Shi**: Conceptualization, Methodology, Writing-Review and Editing, Supervision.

## Funding

This work was funded by Technology Cooperation Project of Jiangsu Province (Grant No. BZ2020022).